\documentclass{article}
  \date{}
\usepackage[utf8]{inputenc}

\usepackage{pifont} 
\usepackage[nottoc,notlot,notlof]{tocbibind}

\usepackage[paper=a4paper,left=30mm,right=20mm,top=25mm,bottom=30mm]{geometry}

\usepackage{hyperref}
\usepackage{tocloft}

\usepackage{etoolbox}

\usepackage{comment}

\usepackage{hyphenat}

\usepackage{natbib}
\usepackage{graphicx}

\usepackage{natbib}
\usepackage{paralist}

\title{Syndromic Surveillance}
 \title{ Continuous Artificial Prediction Markets as a Syndromic Surveillance Technique}
\author{Fatemeh Jahedpari}

\begin{document}

\maketitle

 \section{Syndromic Surveillance}\label{Syndromic Surveilla}\label{GFT}

Appearance of highly virulent viruses warrant early
detection of outbreaks to protect community health.   
The main goal of public health surveillance and more specifically  \emph{`syndromic surveillance systems'}   is early detection of an outbreak in a society using available data sources.  

In this paper, we discuss what are the challenges of   syndromic surveillance systems and how continuous Artificial Prediction Market (c-APM) \citep{jahedpari2017online}  can  effectively be applied to the problem of syndromic surveillance.
 c-APM can effectively be applied to the problem of syndromic surveillance by  analysing each data source with a selection of algorithms and integrating their results according to an adaptive weighting scheme.  
 Section~\ref{introSS} provides an introduction and   explains syndromic surveillance. Then, we discuss  the  syndromic surveillance data sources in Section~\ref{subsec:DataSources}  and  present  some    syndromic surveillance systems in Section~\ref{existingSS}.   The statement of the problem in this field is covered in Section~\ref{ProblemSS}.  After that, we discuss  Google Flu Trends (GFT)  and  GP model~\citep{lampos2015advances},  which is proposed by  Google Flu Trends team to improve GFT engine performance,  in Section~\ref{GFTCase} and~\ref{GPCase} respectively. Also, in these sections, we  evaluate the performance of     c-APM as a syndromic surveillance system. 
 Finally, Section~\ref{ssConclusion} provides the conclusion of this paper.

\section{Introduction}\label{introSS}


 According to the World Health Organisation (WHO)~\citep{whoWebsite}, the United Nations directing and coordinating health authority, \emph{ public health surveillance} is: 
 
 \begin{quotation}
 \noindent The continuous, systematic collection, analysis and interpretation of
 health-related data needed for the planning, implementation, and evaluation of
 public health practice. \end{quotation}

Public health surveillance practice has evolved over time. Although it  was
limited to pen and paper at the beginning of 20th century, it is now facilitated by huge
advances in informatics. Information technology enhancements have changed the
traditional approaches of capturing, storing, sharing and analysing of data and
 resulted efficient and reliable health surveillance techniques~\citep{lombardo2007disease}. 
The main objective and challenge of a health surveillance system is the
 earliest possible detection of a disease outbreak within a society for the purpose of protecting
 community health.

In the past, before the widespread deployment of computers, health surveillance was based on reports received from medical
care centres and laboratories. Although they are very specific\footnote{Specificity: the proportion of people without the disease that a test
finds negative }, they decrease the timeliness and sensitivity\footnote{Sensitivity: the proportion of people with the disease that a test
finds positive} of a surveillance system~\citep{lombardo2007disease},  while
prevention of mortality of infected people for some diseases requires rapid
identification and treatment. Clearly, the earlier a health threat within a
population is detected, the lower the morbidity and the higher number of the saved lives.
Consequently, syndromic surveillance systems have been created to monitor
indirect signals of   disease activity such as call volume to telephone
triage advice lines and over-the-counter drug sales to provide faster detection
\citep{ginsberg2008detecting}. 

Syndromic Surveillance is an alternative   to  the
 traditional  health surveillance system, which   mainly depends  on confirmed diagnoses, and aim  to detect an outbreak as
 early as possible. Syndromic
 surveillance refers to techniques relying on population health indicators which
 are apparent before confirmatory diagnostic tests become available
~\citep{mandl2004implementing}. Syndromic surveillance systems mostly concentrate
 on infectious diseases such as severe acute respiratory syndrome (SARS), anthrax
 and influenza.  In order to decide
 whether an outbreak is evolving,  syndromic surveillance systems monitor the
 quantity of patients with similar syndromes   since indicators of a
 disease appear.

\begin{figure}[ht]
\centering \includegraphics[width=0.9\textwidth]{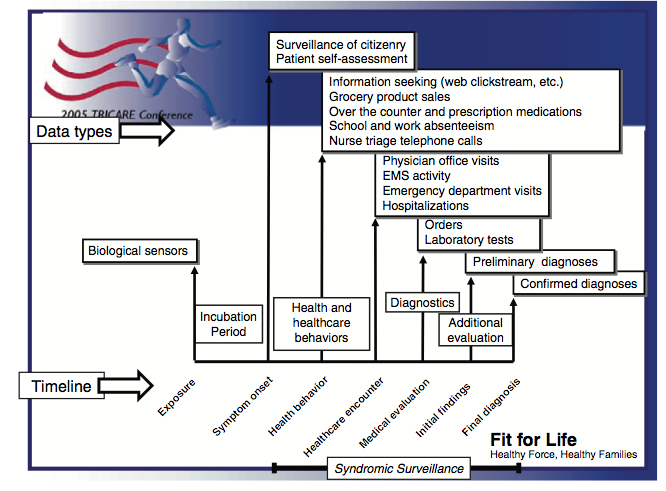}
\caption{Conceptual timeline of pre-diagnosis data types and sources for
syndromic surveillance~\citep{chen2010infectious}.}
\label{fig:DataSourcetimeline}
\end{figure}

Syndromic surveillance aims to exploit information which is not primarily generated 
for the purpose of public health, but can be an indicator of an abnormal health 
event. 
Syndromic surveillance data sources include, but are not limited to, coding of diagnoses
at admission to or discharge from emergency departments, confirmatory
diagnostic cases, medical encounter pre-diagnostic data, absentee
rates at schools and workplaces, over-the-counter pharmacy sales and posts
on social media.  
Each of these  data sources can generate a signal during   disease 
development. 
 Figure~\ref{fig:DataSourcetimeline} shows the timeline of different data sources  to detect an outbreak. 
  The following section describes some of the syndromic surveillance data sources in more details.

\section{Syndromic Surveillance Data Sources}\label{subsec:DataSources}
 
Syndromic surveillance data sources should supply
timely and pre-diagnosis health indicators. Most of this data is  originally
collected for other purposes and now serves a dual
purpose~\citep{chen2010infectious}. 
Syndromic surveillance data sources include:

\begin{enumerate}

\item Chief complaint record: These records include signs and symptoms of
      patient illness from emergency departments (ED) and ambulatory visits  to hospitals.
      These records normally become available on the same day as the patient is
      seen.

\item Over the counter (OTC) sales: since some people may consider visiting a pharmacy
      rather than a physician in their early stage of sickness, these data might
      be more timely. They include detailed information and are available in  near real
      time in electronic format. However, they might be affected by factors such as sales
      promotions, stockpiling of medicines during a season, and product
      placement changes in pharmacies. 

\item School or work absenteeism: Although absenteeism data seems to have good
      timeliness,   their lack of medical detail complicates interpretation~\citep{van2008validation}.

\item Hospital admission records: These data are not sufficiently timely as  
      it might take several days from a patient's first visit  until his/her
      hospitalisation.
      
\item Pre-diagnostic clinical data: These are  indications  by an illness
      before being confirmed via  laboratory tests and include comments
      of health care practitioners, patient encounter information,
      triage nurse calls, 911 calls and ambulance dispatch calls. They are
      relatively timely.

\item International Classification of Disease 9th edition (ICD-9) and
      International Classification of Disease, 9th edition, Clinical
      Modification (ICD-9-CM): These are widely used in many syndromic
      surveillance systems due to their electronic format. They are usually
      generated for billing and insurance reimbursement purposes.
       
\item Laboratory test orders and results: Although laboratory test results are
      very reliable, they lack  timeliness as they usually take a week to be
      completed.
      
\item Emergency Department (ED) diagnostic data: These are regularly available in electronic format but
     takes several days to be prepared.

\item Internet and open source information: These contain a huge source of health
      information and can be obtained via discussion forums, social media,
      government websites,  news outlets, blogs, discussion sites, individual
      search queries, web crawling, use of click stream data, mass media and
      news report. 
 
      For example,  some approaches have applied data mining techniques to   
      
      \begin{itemize}
      \item  Search engine logs
      
       \citep{eysenbach2006infodemiology},\citep{polgreen2008using},\citep{eysenbach2009infodemiology},~\citep{ginsberg2009detecting}, \citep{lampos2010tracking} and \citep{lampos2015advances}

      \item Twitter
      
       \citep{culotta2010detecting},\citep{achrekar2011predicting},\citep{Signorini2011},\citep{culotta2013lightweight} and \citep{paul2014twitter}  
      
      \item  News articles 
      
        \citep{reilly1968indications},~\citep{grishman2002information},\citep{mawudeku2006global},\citep{brownstein2008surveillance},~\citep{ collier2008biocaster} and \citep{linge2009internet}
      
      \item      
      
      Web browsing patterns 
      
      \citep{johnson2004analysis}) and  blogs (\citep{corley2010text}) 
      \end{itemize}

    \end{enumerate}

\begin{figure}[ht]
\centering
\includegraphics[width=1\textwidth]{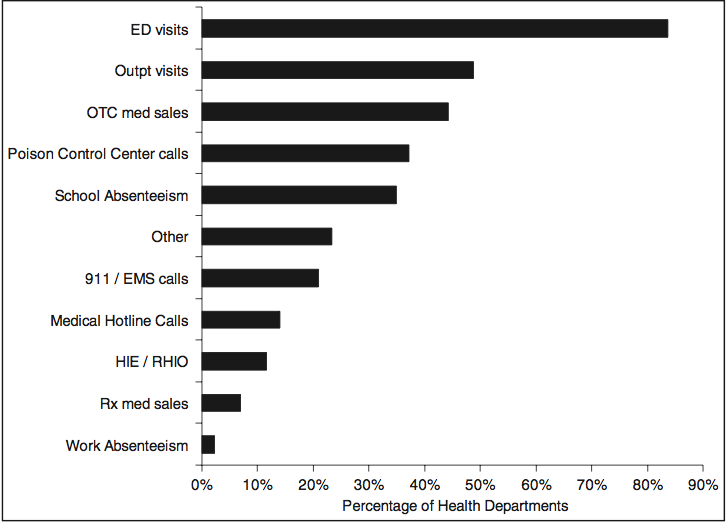}
\caption{Distribution of different data source usage in existing syndromic
surveillance systems in the USA~\citep{Buehler2008}.}
\label{fig:DataSourceSurvey}
\end{figure}

Figure~\ref{fig:DataSourceSurvey},
graphs the popularity of various data sources in existing syndromic surveillance
systems in the USA. As can be seen from the figure, while emergency department visit
reports are widely used in such systems, work absenteeism is the least popular
source.

 
\section{Existing Syndromic Surveillance Systems}\label{existingSS}

In recent years, a number of syndromic surveillance approaches have been
proposed. Roughly 100 syndromic surveillance systems were deployed   in the USA done 
by 2003~\citep{buehler2003syndromic}. Although they share similar goals, they are
different in their system architecture, information processing, analysis
algorithms, disease focus, and cover different geographic locations.~\cite{chen2010infectious} summarises the main international and USA local,
state and national syndromic surveillance systems. In Europe, an inventory of syndromic
surveillance systems  is delivered through a new Public Health Action
Programme called Triple-S\footnote{ http://www.syndromicsurveillance.eu/ (Retrieved  Oct 4, 2015). }
(Syndromic Surveillance Survey, Assessment towards Guidelines for Europe). 

 The following two sections survey some of the major existing syndromic surveillance systems around the globe.
Based on the utilised data sources, we divide the existing syndromic surveillance systems into two categories of 
\begin{inparaenum}[i)]
\item traditional syndromic surveillance systems, described in Section~\ref{tss}  and  
\item modern syndromic surveillance systems, described in Section~\ref{modernSS}.  
\end{inparaenum}

\subsection{Traditional Syndromic Surveillance Systems}\label{tss}

We refer to syndromic surveillance systems that do not utilise social media and internet based data  as traditional syndromic surveillance. Some of them are
listed below: 

\begin{enumerate}

\item Early Notification of Community-based Epidemics (ESSENCE)
\citep{lewis2002disease} is a syndromic surveillance system in the Washington D.C. area, undertaken by
Department of Defense with the primary goal of early detection disease outbreak due
to bioterrorism attacks. 

\item Real time Outbreak and
Disease Surveillance (RODS)~\citep{tsui2003technical} is a public health
surveillance system, in operation in western Pennsylvania since 1999, developed
at the RODS laboratory of the Center for Biomedical Informatics at the
University of Pittsburgh. 

\item Composite Occupational Health and Operational Risk
Tracking (COHORT)~\citep{reichard2004cohort} delivers real-time surveillance of
the medical care of specified groups of military employees worldwide.

 \item Syndromic Surveillance Information Collection (SSIC)
has been developed by the association of the Clinical Information Research Group at
the University of Washington and Public Health-Seattle and King County
\citep{lober2003syndromic}. 

\item  Infectious Disease
Surveillance Information System (ISIS)~\citep{widdowson2003automated} is an
automated outbreak detection system for all types of pathogens in the
Netherlands.

\item Early Aberration Reporting System (EARS) is developed by Center for Disease
Control (CDC)~\citep{hutwagner2003bioterrorism} and enables national, state and
local health departments to analyse public health surveillance data using a collection
of anomaly detection methods.

\item Japan National Institute of Infectious Diseases
(NIID)~\citep{ohkusa2005experimental} has developed syndromic surveillance system to analyse
over the counter  sales data, outpatient visits, and ambulance transfer data in Tokyo.


\end{enumerate}

 We now provide a detailed description of two of the popular traditional syndromic surveillance system, namely BioSense and PHE ReSST.

\subsubsection*{BioSense}\label{BioSense}

BioSense\footnote{http://www.cdc.gov/BioSense/ (Retrieved  March 4, 2015). } is a syndromic surveillance
system in the United State which is part of CDC's Public Health Information Network
framework. By monitoring the size, location and rate of spread of an outbreak,
it detects an outbreak at the local, state and national levels. It monitors
seasonal trends for influenza and other disease indicators.
BioSense concentrates on syndrome categories including fever, respiratory,
gastrointestinal illness (GI), hemorrhagic illness, localised cutaneous lesion,
lymphadenitis, neurologic, rash, severe illness and death, specific infection,
and botulism.

BioSense collects and shares information on emergency department visits,
hospitalisations, clinical laboratory test orders, over-the-counter (OTC) drug
sales and other health related data from multiple sources, including the
Department of Veterans Affairs (VA), the Department of Defense (DoD), and
civilian hospitals from around the USA.  
BioSense uses multiple analysing methods such as CUSUM~\citep{page1954continuous}, EWMA~\citep{EWMA} and SMART~\citep{kleinman2004generalized}.

\subsubsection*{PHE ReSST } \label{NHSDirectSSS}
 
The Public Health England (PHE)\footnote{http://www.hpa.org.uk/  (Retrieved  March 4, 2015).} Real-time
Syndromic Surveillance Team (ReSST) generates regular syndromic surveillance
reports by collaborating with numerous national syndromic surveillance systems
including the  NHS Direct syndromic surveillance system. The NHS Direct syndromic
surveillance system monitors the nurse-led telephone helpline data collected
electronically by NHS Direct sites and generates alarms when call numbers are
considerably higher than preceding years, after considering holiday and seasonal
effects. It has the potential to detect large scale events, but is less likely to
detect smaller and localised outbreaks~\citep{doroshenko2005evaluation}. In
addition, ReSST obtains data from GP In-Hours and GP Out-of-Hours syndromic
surveillance systems which monitor daily consultations for a range of clinical
syndromic indicators and community-based morbidity, recorded by GP practices
inside and outside of routine surgery opening times, respectively. 

\subsection{Internet-Based Syndromic Surveillance Systems } \label{modernSS}

There are other real-time disease event detection systems which employ different
approaches from the systems discussed in Section~\ref{tss}.  They monitor online media from global sources, instead of monitoring
disease cases reported by health related organisations such as hospitals and
clinics. These ``systems are
built on top of open sources, exemplifying an idea of open development for
public health informatics applications''~\citep{chen2010infectious}. 
Though the modern systems are
faster than traditional syndromic surveillance systems in detecting an anomaly
in public health~\citep{Signorini2011, ginsberg2008detecting}, they are vulnerable to a high rate of false positives
in case of an unusual event within a population~\citep{ginsberg2008detecting}. 
This section describes some of the well known modern syndromic surveillance systems.

\subsubsection*{Google Flu Trends}

Google Flu Trends\footnote{http:// www.google.com/flutrends/  (Retrieved  March 7, 2014).}, established
by Google, is a Web-based tool for near real-time detection of regional
outbreaks of influenza~\citep{ginsberg2008detecting}. It monitors and analyses
health-care seeking behaviour in the form of queries to its online search engine.
According to ~\citet{ carneiro2009google} ``all the people
searching for influenza-related topics are not ill, but trends emerge when all
influenza-related searches are added together''; Consequently, there is a close
relationship between the number of people searching for influenza-related
topics and those who have influenza symptoms. Section~\ref{GFTCase} provides  more information about Google Flu Trends.

\subsubsection*{Argus}

The Argus system is a web-based global biosurveillance system   designed to
report and track the development of biological events threatening human, plant
and animal health globally, excluding the USA~\citep{united2009one}. It is developed
at Georgetown University and funded by the United States Government.

It automatically collects local and native language internet media reports
including blogs and official sources such as World Health Organisation (WHO) and
World Organisation for Animal Health (OIE) and infers their importance according
to keywords appropriate to infectious disease surveillance
~\citep{nelson2010event}. It relies on a human team of multilingual data analysts to assess the
relations between the online media and presence of adverse health events
\citep{chen2010infectious}.  In particular, the data analysts monitor several thousand Internet sources    daily.  Then, six time in each day, they use Boolean keyword searching and Bayesian model tools~\citep{mccallum1998comparison} to select    relevant media reports~\citep{nelson2010event}. Based on the selected media reports, they write their own report and post them on a secure Internet portal to be accesses with   Argus users.

Since its operation in July 2000, ``it has logged more than 30,000 biological
events involving pathogens such as avian influenza, the Ebola virus, cholera, and
other unusual pathogens that have caused varying states of social disruption
throughout the world"~\citep{collaborationbrief}.

\subsubsection*{GermTrax}

GermTrax\footnote {http://www.germtrax.com/  (Retrieved  Oct 4, 2015).} is a freely accessible website which
gathers sickness and disease data from people worldwide and exhibits trends
through an interactive map. More specifically, GermTrax is a collaborative disease tracking system which   primarily relies on
reports filled by ordinary people who are sick.
This system collects information through user 
 personal updates on social media websites such as Facebook and Twitter.
Then, the system   saves user geo-location
data,  while the users connect their social media accounts with the site. 
According to
their website, GermTrax can help people by informing them of  places where  they might get sick
and help health experts to discover large-scale sickness trends. Since it principally
relies on disease reports from ordinary people, it is suitable for
non-specific conditions such as colds and flu~\citep{lan2012picture}.

\subsubsection*{Health Map}

Health Map\footnote{http://www.healthmap.org   (Retrieved  Oct 4, 2015).} is a multi stream real-time
surveillance system and freely accessible. It monitors online information in
order to obtain a comprehensive view of current infectious disease outbreaks   globally. It observes,   filters,
visualises, and distributes online information about emerging infectious
diseases for the benefit of diverse audience from public health officials to
international tourists~\citep{lemon2007global}. 
Health Map gathers
reports from 14 sources, which in turn embody information from over 20,000 web
sites every hour. Information is obtained automatically through screen scraping,
natural language interpretation, text mining, and
parsing~\citep{brownstein2008surveillance}.  More specifically, 
Health Map use multiple web based data sources including online news sources, expert-curated discussion, and validated official reports from organisations such as the World
Health Organisation (WHO\footnote{http://www.who.int/en/(Retrieved  Oct 4, 2015).}). 
Then,  the alerts are classified by location and disease  using  automated text processing algorithms. Next,   the system overlays the alerts on an interactive geographic map. According to~\cite{freifeld2008healthmap} ``The filtering and visualization features of HealthMap thus serve to bring structure to an otherwise overwhelming amount of information, enabling the user to quickly and easily see those elements pertinent to her area of interest''.

\section{Statement of the Problem}\label{ProblemSS}

 While traditional syndromic surveillance systems can detect an outbreak with
high accuracy,
 they suffer from slow response. For example, Centers for Disease Control and
 Prevention (CDC) publishes USA national and regional data typically with a 1-2 week
 reporting lag   using outpatient reporting and virological test results provided by laboratories nationally
\citep{culotta2010detecting,culotta2013lightweight, ginsberg2008detecting}. 
 Therefore, such systems  cannot predict an outbreak, but only can detect them after  the  onset.

 On the other hand, modern syndromic surveillance systems monitor  online media from global sources. 
  Such
 modern syndromic surveillance systems resort to internet
 based data such as search engine queries, health news, and people posts on
 social networks to predict an outbreak earlier~\citep{Signorini2011,
 carneiro2009google,corley2010text}. While some of them claim
 that they could achieve high accuracy, the rate of false alarms is unknown. 
\cite{ginsberg2008detecting} state, regarding Google Flu 
 Trends, that
 ``Despite strong historical correlations, our system remains susceptible to
 false alerts caused by a sudden increase in ILI-related queries. An unusual
 event, such as a drug recall for a popular cold or flu remedy, could cause such
 a false alert".   Therefore, an issue with  internet based data sources  is that  their data quality fluctuates  over time. 

  Moreover,  most of these modern syndromic surveillance
 systems rely on  one type of internet based data sources and disregard the advantage of
 other type of data sources, which are discussed in Section
~\ref{subsec:DataSources} (page~\pageref{subsec:DataSources}). Consequently, they are only suitable for  places
 where their source data is sufficiently available. For example, Twitter  based
 systems cannot have a high accuracy for places where using twitter is not  very
 common, if accessible.  In addition, the quality and availability of data sources may change over time. For instance,  Twitter may lose its popularity or become inaccessible in    a place.
 Hence, integrating available data sources according to an adaptive weighting scheme over time seems necessary.

The other area that has received attention in the syndromic surveillance literature is the topic of alternative analysis algorithms for a given data sources.  Given that the quality of data sources change  over  time, and the most suitable  algorithm for a given data source is not known \emph{a priori}, a reasonable response is to consider analysing each data source with a variety of algorithms and integrate their results. 

Against this background, we believe, based on plentiful available data sources and analysis  techniques,  a state of the art syndromic surveillance
  mechanism should:

 \begin{enumerate} 
    
\item Perform as an ensemble to combine various analysis  algorithms with the objective of increasing   syndromic surveillance system performance. There are many different techniques
      with different strengths and weaknesses. An ensemble which utilises a
      combination of them seems likely to be able provide higher performance than systems which
      are depended on only one technique.  \label{r1}

\item Extract information which resides in different data
      sources. 
       In addition to
      obtaining information, it should be capable of integrating them according to their  relevance and  varying quality.   \label{r2}

\item Be flexible to  changes in composition  of algorithms and data sources 
      over time as any of them might be deleted, temporarily unavailable, or
      added to the system at any time.
      \label{r3} 
      
\item Be able to adapt to its corresponding monitored population behaviour and
      habits.  For example, if  people of a particular region  are more prone  to tweet their feeling in social media such as Twitter than searching for a solution  using  online search engines, then a syndromic surveillance system should weight  twitter results  higher  than a search engine queries  in that particular region.
     
       \label{r4} 
      
\item Be able to adapt to the \emph{changes} of its corresponding
      population  behaviour. For example, if twitter
      become more popular in a place and people start tweeting their sickness symptoms
      earlier, rather than visiting a physician, the system must give more attention and
      weight to twitter than previously.  \label{r5}

 \item Minimise the effect of misleading factors and noise 
      such  as advertisement, promotions, and holidays on different data
      sources and, consequently, diminish the rate false positives.   \label{r6}

  \end{enumerate}

 \citet{jahedpari2017online}  proposed Continuous Artificial Prediction Market (c-APM), which utilizes the concept of prediction markets  in which the traders are modeled as intelligent agents. The model can be used as a machine learning ensemble by integrating different data sources and techniques.
   
  In here, we suggest that c-APM can be used as a syndromic surveillance technique as it  fulfills the aforementioned requirements as we discuss below:

    \begin{enumerate}   [ 1)]
  
\item  c-APM  can behave as an ensemble method by including
    numerous agents, each having different analysis  algorithms. 
    
    \item Prediction markets are specially designed for the purpose of
      information aggregation~\citep{perols2009information}. 
c-APM adapt the prediction markets' concepts and incentives it participating agents to  share their private information through market mechanism,   hence make accurate prediction. In addition, c-APM dynamically  weights the prediction of different agents  according to their varying quality. 
  
  \item  In c-APM, market and other agents operate independently and hence absence or presence of an agent does not impact
      the system considerably. Therefore, if one of the existing data sources becomes unavailable for any reasons, c-APM can simply
      respond   to the issue. 
      If a new data source or a model is discovered,  c-APM can simply create an agent to access that data source or model to participate in the market and share its knowledge. 

      \item In c-APM,  
       the agents can be trained in the market
      using historical data of that place and, consequently  will be adapted to behaviour of
      people in that place. 

      \item   c-APM can respond  to the  changes  of its corresponding
      population  behaviour  since its agents keep learning  and their weights keep changing according to their current performance in  each
      market. 

      \item c-APM can minimise the effect of misleading factors and noise 
       by fusing various data sources and models using an adoptable scheme. 
      
        \end{enumerate}

  In the following sections, we use two well-known models of  (i) Google Flu Trends, and  (ii) the latest improvement of  Google Flu Trends model, named as GP~\citep{lampos2015advances},  as our case study and we show  how c-APM can improve upon their performance.

 \section{Google Flu Trends Case Study}\label{GFTCase}

 Google Flu Trends (GFT) was launched  by Google in 2008 to alert health professionals to outbreaks early by indicating when and where influenza is striking in real time using aggregate web searches.  GFT   publishes  flu predictions (ILI rate) for more than 25 countries.
 Google Flu Trends is   typically more immediate, up to  2 weeks ahead of traditional methods such as the CDC's  official reports.  The basic idea behind GFT is that when people get sick, they turn to the Web for information.

   Google Flu Trends  algorithms recognise  a small subgroup of the millions of search engine query terms that deliver the maximum correlation with the CDC published ILI rate. Then  a subset of these queries which fit the historical CDC ILI rate   data   most accurately  are chosen.  
Finally, univariate linear regression model is trained to be used in predicting future ILI rate using each day queries. 
 According to~\citet{copeland2013google}  the challenge of their approach is the varying volumes of a particular query   over time. 
 For instance, during the holiday season, more people search for `gift' than at any other period. Similarly, overall usage of Google search varies throughout the year and is growing over time. 
 GFT   used the official CDC data only in the initial training and    did not use it   to  re-train  its model  regularly\footnote{\url{http://googleresearch.blogspot.ae/2014/10/google-flu-trends-gets-brand-new-engine.html} (Retrieved  Oct 4, 2015).}.

  The early Google paper indicated that the Google Flu Trends predictions were 97\% accurate comparing with CDC data~\citep{ginsberg2009detecting}. However, in 2013,~\cite{olson2013reassessing} and~\cite{butler2013google}   reported that GFT was predicting more than double that of CDC published. Later in 2014,~\cite{lazer2014parable} stated     that GFT has been   overestimating flu occurrence for most  weeks after August 2011 and by a very large margin in the 2011-2012 flu season. 
    He continued stating GFT can achieve better performance by combining its prediction with other near realtime health data such as lagged CDC data.
Also, Google Flu Trend team announced\footnote{http://blog.google.org/2013/10/flu-trends-updates-model-to-help.html (Retrieved  Oct 4, 2015).}
 \begin{quotation}  
 ``We found that heightened media coverage on the severity of the flu season resulted in an extended period in which users were searching for terms we've identified as correlated with flu levels. In early 2013, we saw more flu-related searches in the US than ever before.''
  
  \end{quotation}  

GFT subsequently  updated the model    in response to concerns about accuracy.  
In 9th August 2015, GFT stopped publishing flu predictions without formally presenting any reasons. However, GFT historical prediction are still available for download.

 \subsection{Comparison of c-APM  and GFT}\label{ComparewithGFT}

In this section, we use c-APM as a syndromic surveillance system and compare the performance of c-APM and Google Flue Trend.

\subsubsection{Experimental  Setup}\label{settingSSChapeterGFT}

In these experiments, c-APM predicts  the disease activity level of influenza-like illnesses (ILI) in a given week in the whole of the USA using publicly available data sources.  The data  used here contains more than 100 real data sources covering the period 4th January 2004 (when GFT provides data for most of USA states and cities) to 9th August 2015 (when GFT stopped publishing their results online), from the two data sources of Google Flu Trends (GFT) and Centers for Disease Control and Prevention (CDC). 

\subsubsection*{Data Sources}
 In these experiments, we   use weekly Google Flu Prediction for different areas of the United States including states, cities and regions\footnote{  
 This data can be accessed from \url{https://www.google.org/flutrends/about}. (Retrieved  Oct 4, 2015).}, for which GFT data is available since  January 2004. 

In here, we use the calendar  definition of year where a year starts on 1st January and finishes on 31st  December. 
 
  The CDC Influenza Division produces a weekly report on influenza-like illness\footnote{ILI is defined as fever (temperature of 100$^{\circ}$F [37.8$^{\circ}$C] or greater)  and a cough and/or a sore throat without a known cause other than influenza (\url{http://www.cdc.gov/flu/weekly/overview.htm}) (Retrieved  Oct 4, 2015).} activity in the USA\footnote{  
 This data can be accessed from \url{http://gis.cdc.gov/grasp/fluview/fluportaldashboard.html}. (Retrieved  Oct 4, 2015).}.  
We   use CDC statistics  including:
\begin{inparaenum}[i)]
\item    ILI rate disaggregated by age groups  
(0-4 years, 5-24 years, 25-64 years, and older than 65 years), 
\item USA national ILI rate, 
\item total number of patients and
\item  total number of outpatient healthcare providers in  U.S. Outpatient Influenza-like Illness Surveillance Network (ILI network). 
 \end{inparaenum}
Since CDC reports ILI rates with a two-week time lag, we use CDC data of two weeks earlier  for each week of the experimentation period.  In this way, we can align CDC data with the other  data sources used in these experiments.

\subsubsection*{Models}
 We use different machine learning models in  R's caret package  (version 6.0-37), which are capable of performing regression.   Table~\ref{RModelsforGftComparison} presents the models we use in this experiment.  Model parameters are set to their default values.
  

\begin{table}[]
\centering
\begin{tabular}{|l|l|}
\hline
\textbf{Model Full Name}                & \textbf{Mdoel Short Name} \\ \hline
Bagged CART                             & treebag                   \\ \hline
Conditional Inference Random Forest     & cforest                   \\ \hline
Random Forest                           & rf                        \\ \hline
Multi-Layer Perceptron                  & mlp                       \\ \hline
Model Averaged Neural Network           & avNNet                    \\ \hline
Boosted Generalized Linear Model        & glmboost                  \\ \hline
Boosted Tree Linear Regression          & blackboost                \\ \hline
Linear Regression                       & lm                        \\ \hline
Radial Basis Function Network           & rbf                       \\ \hline
Gaussian Process                        & gaussprLinear             \\ \hline
CART                                    & rpart                     \\ \hline
Generalized Linear Model                & glm                       \\ \hline
k-Nearest Neighbors                     & knn                       \\ \hline
Gaussian Process with Polynomial Kernel & gaussprPoly               \\ \hline
Multivariate Adaptive Regression Spline & earth                     \\ \hline
Self-Organizing Map                     & bdk                       \\ \hline
\end{tabular}
\caption{ R's caret package models. c-APM  instantiates one participant for each of these   models. }
\label{RModelsforGftComparison}
\end{table}

 
 \subsubsection*{ Experiment Settings}
 We constructed an c-APM 
in which every agent has a unique analysis model corresponding to one of the models listed in Table~\ref{RModelsforGftComparison}. The data source for each agent is the entire data set. 
 All agents   use  Q-learning trading strategy, which is proposed in  \cite{jahedpari2017online}.   The results  are based on one run only, as they   are deterministic.   All  c-APM parameters are set to their default parameters (see \citep{jahedpari2016artificial}). Hence: 
\begin{enumerate}[ i)]
\item  The number of rounds is set to 2, 
\item $MaxRPT$ and  $MinRPT$  is set to $90\%$, in the first round, and  
\item  $MinRPT$ and $MaxRPT$ are set to $0.01\%$ and $1\%$ respectively,  in the second round.
 \end{enumerate}

We measure the performance of c-APM by comparing the prediction of c-APM against the ground truth, which is the weekly ILI rate published by CDC.   We use Mean Absolute Error (MAE), which is a common  measure, in this literature.  
 
 \subsubsection{Experimental Results}\label{resultswithGFT}

 \begin{figure}[!t]
 
                \includegraphics[width=0.95\columnwidth]{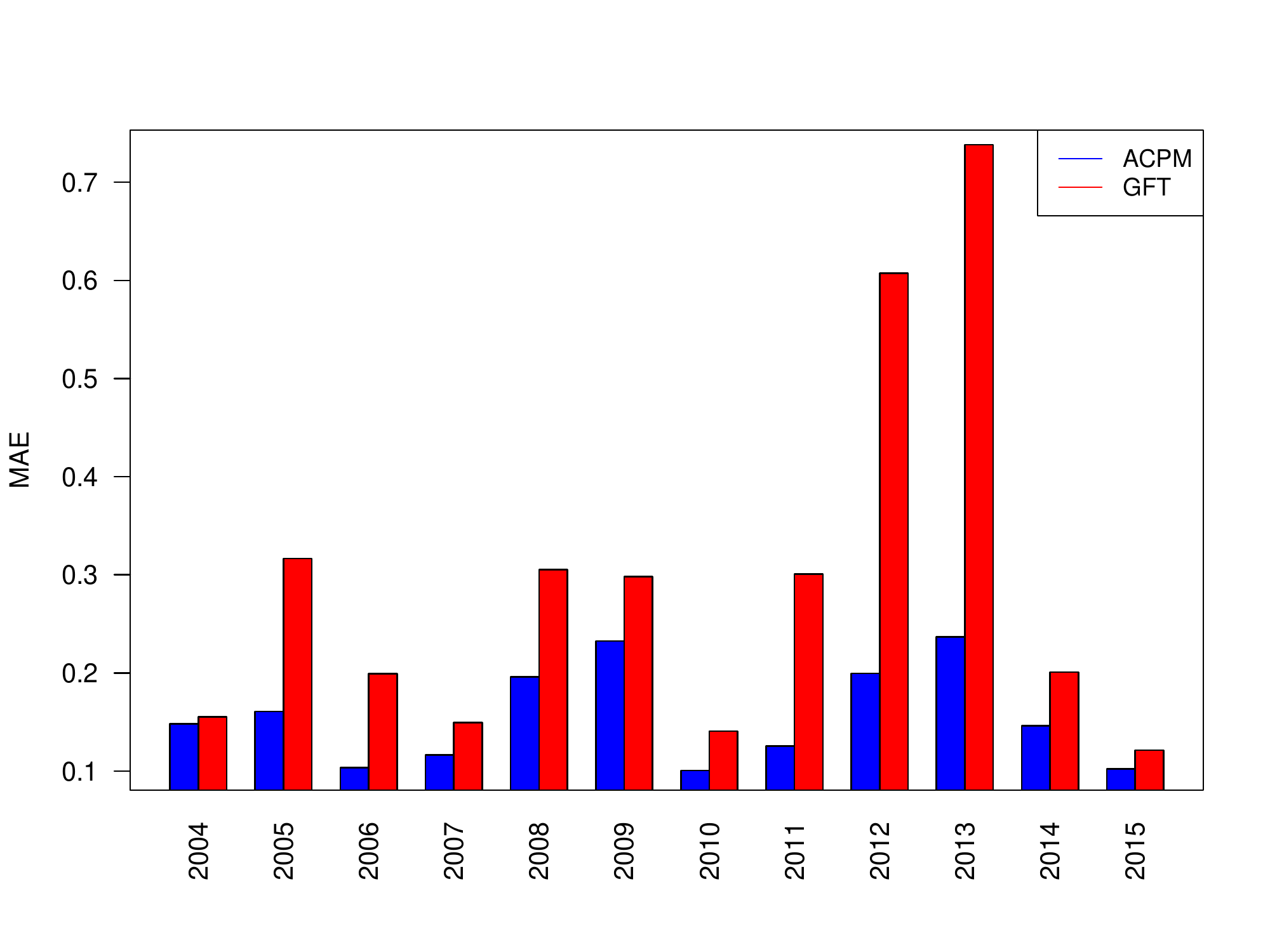}
                \caption{Comparing the Performance of c-APM and GFT for different periods using Mean Absolute Error (MAE). }
                \label{GFT-ErrorForeachPeriod}
             
\end{figure}

\begin{figure}[t!]
\includegraphics[width=0.95\columnwidth]{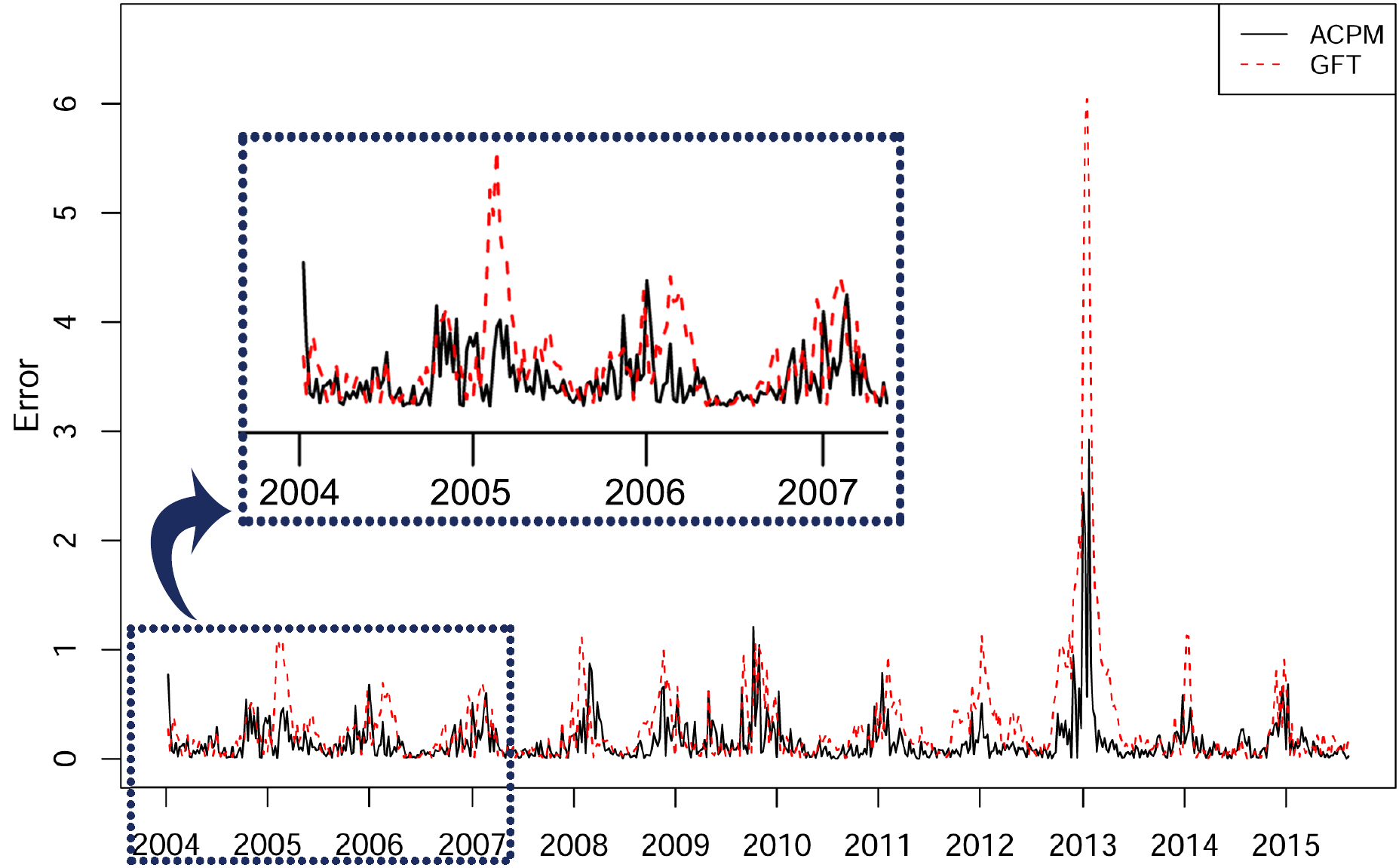}
\caption{c-APM and GFT error in predicting ILI rate from 2004 to 2015}
\label{errorCompareGFTall}
\end{figure}

 In this section, we compare the performance of c-APM and Google Flu Trends. 
 Figure~\ref{GFT-ErrorForeachPeriod} and Figure~\ref{errorCompareGFTall} compare the error of c-APM and Google Flu Trend for the period between $2004$ to $2015$.    
 
 As  Figure~\ref{GFT-ErrorForeachPeriod} shows,  c-APM  typically has a lower, sometimes much lower, MAE to that of Google Flu Trends  in each year.   
 Though this difference is relatively small in some years like 2004 and 2007, it is relatively large in most years and very large between 2011  and 2013. 
 Table~\ref{PvalueCompareGFT} shows the exact MAE value of c-APM and GFT in addition to  t-test  p-values. 
The null hypothesis is that the two accuracies compared are not significantly different. Therefore, within a tolerance $\alpha = 0.05$, when p-value~$<0.05$, c-APM is significantly better than GFT.  As the table shows     the results are  highly significant in  most years and also during the entire period  of 2004-2015 (p-value $ = 2.34E-17$). 

  Figure~\ref{errorCompareGFTall} shows that c-APM   performs poorly for the first few  markets which we attribute to the learning period. However, after several   markets, c-APM achieves higher performance than Google Flu Trend in most weeks.     c-APM uses CDC data    as one of its data sources, and since CDC report the data with two weeks time lags, c-APM uses the CDC data of the previous two weeks. This explains the existence of two weeks time lag between c-APM and GFT error in some periods such as early 2008 and late 2012.

 \begin{table}[]
\centering

\begin{tabular}{|c|c|c|c|}
\hline
\textbf{Periods} & \textbf{\begin{tabular}[c]{@{}c@{}}c-APM \\ MAE $\times 10^2$ \end{tabular}} & \textbf{\begin{tabular}[c]{@{}c@{}}GFT \\ MAE $\times 10^2$ \end{tabular}} & \textbf{P-value} \\ \hline
2004             & 0.148                                                        & 0.155                                                       & 3.73E-01         \\ \hline
2005             & 0.161                                                        & 0.317                                                       & 5.82E-05         \\ \hline
2006             & 0.104                                                        & 0.199                                                       & 7.89E-04         \\ \hline
2007             & 0.117                                                        & 0.150                                                       & 4.42E-02         \\ \hline
2008             & 0.196                                                        & 0.305                                                       & 4.15E-03         \\ \hline
2009             & 0.233                                                        & 0.298                                                       & 5.66E-02         \\ \hline
2010             & 0.101                                                        & 0.141                                                       & 2.18E-02         \\ \hline
2011             & 0.126                                                        & 0.301                                                       & 2.33E-08         \\ \hline
2012             & 0.200                                                        & 0.607                                                       & 3.12E-09         \\ \hline
2013             & 0.237                                                        & 0.738                                                       & 4.64E-04         \\ \hline
2014             & 0.146                                                        & 0.201                                                       & 4.85E-02         \\ \hline
2015             & 0.102                                                        & 0.121                                                       & 1.57E-01         \\ \hline
2004-2015        & 0.158                                                        & 0.301                                                       & 2.34E-17         \\ \hline
\end{tabular}
\caption{Performance of c-APM and GFT in predicting ILI rate using Mean Absolute Error (MAE)  and p-values of paired t-test.  }
\label{PvalueCompareGFT}
\end{table}

  \section{GP Case Study}\label{GPCase}
 
 ~\citet{lampos2015advances}  published a paper in Nature Scientific Reports on 3rd August 2015 proposing a  new model, called `GP'.
 Their model includes three improvements to the original  Google Flu Trend. Firstly,  they expand   and re-weight the set of queries which are originally used by GFT.  
  Then, they expand this improvement by using a nonlinear regression framework based on a Gaussian Process (GP) to  investigate nonlinear relationship between query fractions and the ground truth (CDC ILI rate). 
   Finally, they utilise  time series structure. More specifically, they use ARMAX model~\citep{JSSv027i03}  to find a relationship between previously available data and the current one.
  
  They perform an evaluation using five consecutive  influenza seasons, as defined by CDC,   from 2008 to 2013.
  Based on their experiments, they conclude that  GP approach performs better than   GFT and  a well established model, namely Elastic Net. They also mentioned that  2009-10 flu season    is a unique flu period since  during the peak  of that flu season,  GFT over-predicted the ILI rate, while   GP and Elastic Net underestimated the ILI rate.

\subsection{Comparison of c-APM  and GP}\label{ComparewithGP}

This section compares the performance of c-APM and the model proposed by~\cite{lampos2015advances}, known as the `GP' model. We contacted the author and received their exact prediction for each experimented period to use in our experiments.

\subsubsection{Experimental Setup}\label{settingSSChapeterGP}

 All settings are similar to the settings covered in Section~\ref{settingSSChapeterGFT}  (page~\pageref{settingSSChapeterGFT}), except the part that  c-APM includes on additional agent which   uses GP prediction as its data source. The agent uses  a simple   algorithm which gives the prediction equal to the receiving data, hence no analysis is performed by the agent on that data.  
 
In these experiments, we follow the same evaluation format  as the work by~\cite{lampos2015advances}, therefore  we compare the performance of c-APM and GP in the flu seasons  2008 to 2013 as defined by CDC. These flu seasons include different numbers of weeks (see Table~\ref{GpTable}).

\subsubsection{Experimental Results}\label{resultswithGP}
   \begin{figure}[!t]
 
                \includegraphics[width=0.95\columnwidth]{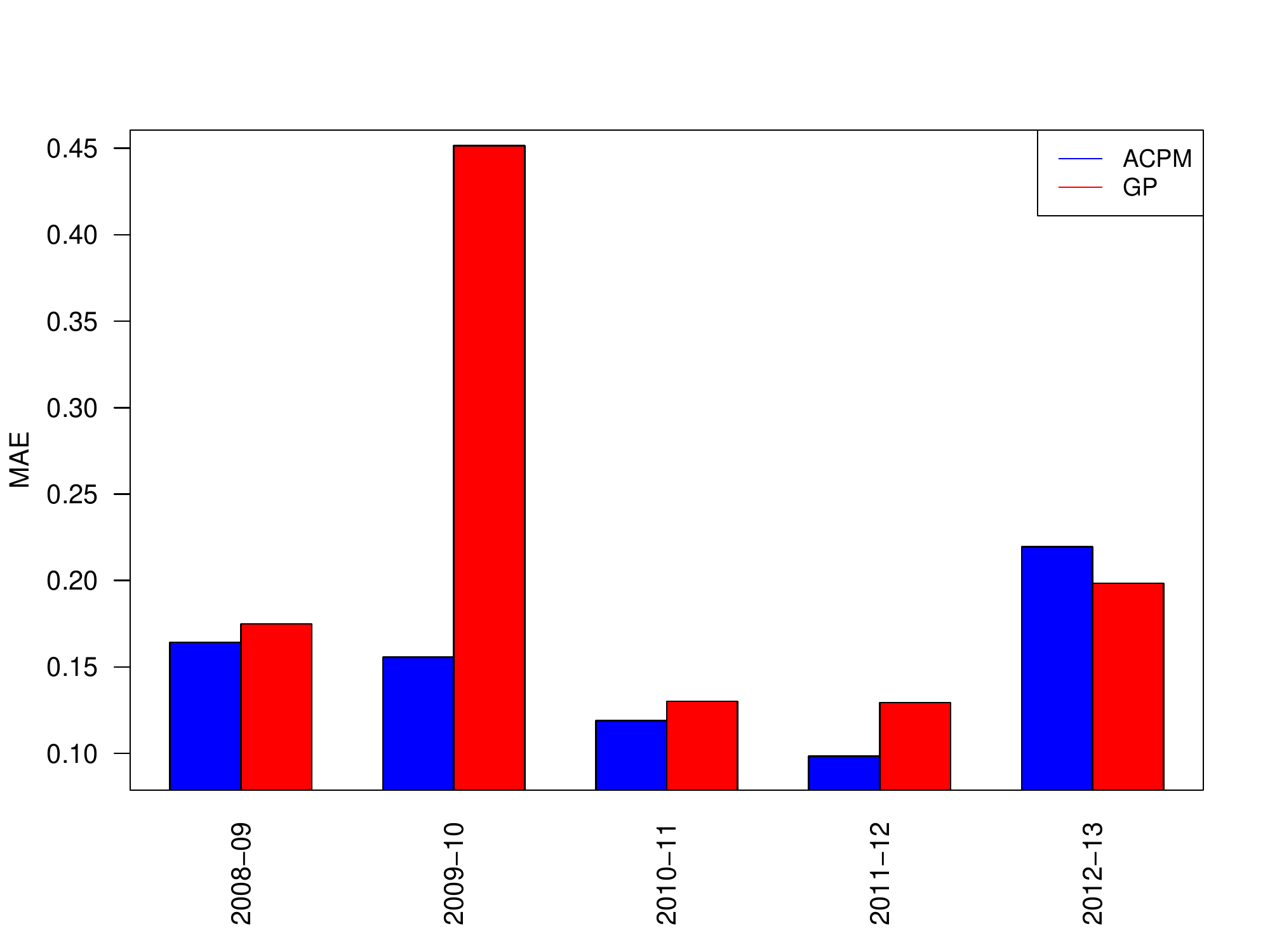}
                \caption{Comparing the Performance of c-APM and GP for different periods using Mean Absolute Error (MAE).}
                \label{GP-ErrorForeachPeriod}
                
\end{figure}

  \begin{figure}[!t]
 
                \includegraphics[width=0.95\columnwidth]{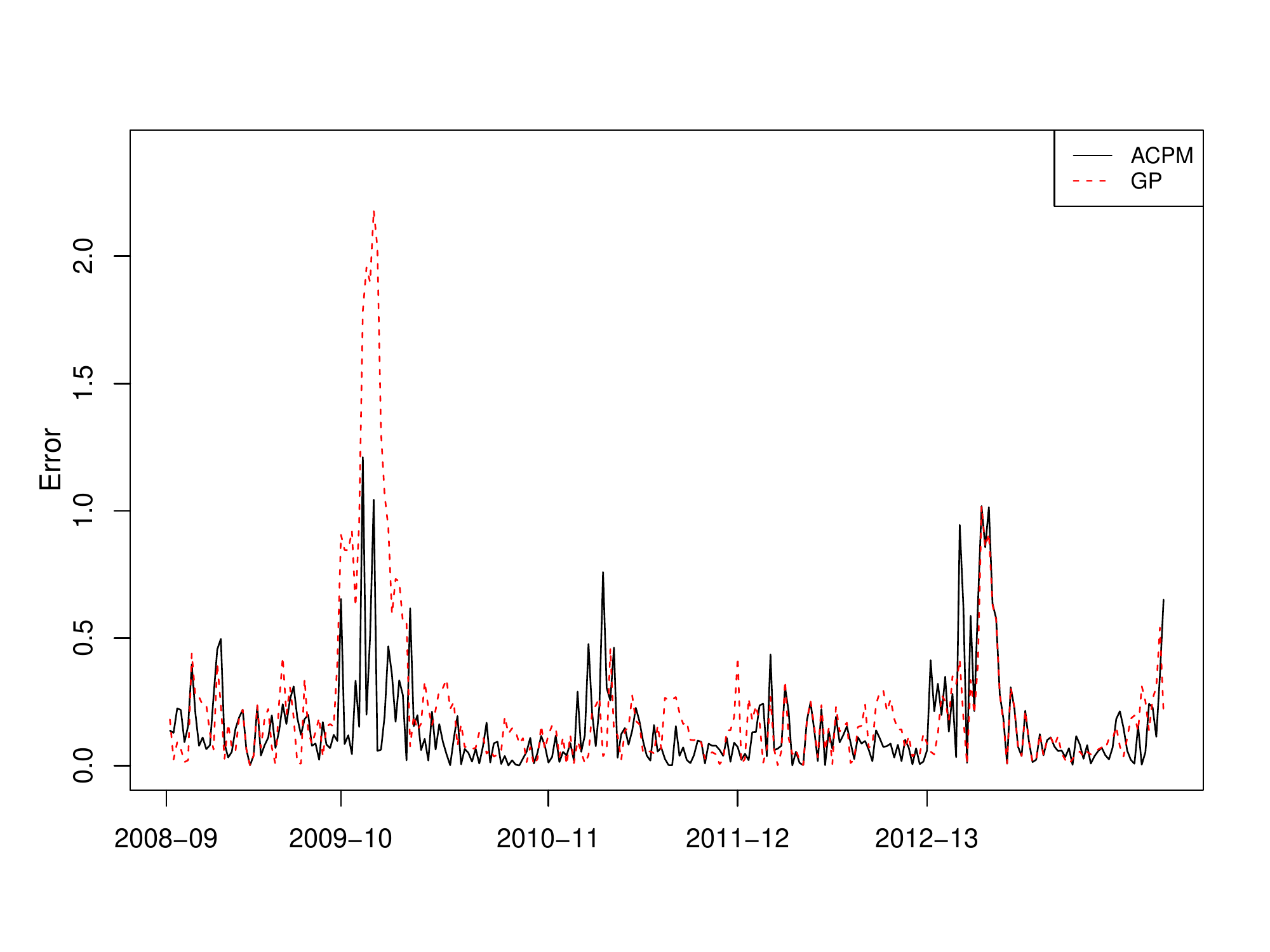}
                \caption{c-APM and GP error in predicting ILI rate from 2004 to 2015 }
                \label{GP2008-2013}
                
\end{figure}

Figures~\ref{GP-ErrorForeachPeriod}  and Table~\ref{GpTable} compare  the performance of c-APM and GP for different influenza seasons  between 2008 and 2013.   Figure~\ref{GP2008-2013} compares the error of c-APM and GP in each week  of the entire period.  In Table~\ref{GpTable}, the first column     shows the experimented influenza seasons and the second  column presents  the number of weeks in each season. 
 The third and the fourth columns show  the  Mean Absolute Error (MAE) of c-APM and GP respectively. The last column shows p-values for the  paired t-tests comparing the error of c-APM and GP.  
 

 As Table~\ref{GpTable} and Figure~\ref{GP-ErrorForeachPeriod}  show c-APM outperforms GP in most years except 2012-2013, where c-APM achieves MAE of $0.220$ and GP achieves MAE of $0.198$.  As shown by Figure~\ref{GP2008-2013}, this is  mainly because of lower performance of c-APM  compared to GP in early weeks of 2012-13 flu seasons.  A few weeks earlier than that,  in the late   2011-12 flu season,  GP performs worse   than c-APM  which infers that the GP agent    performs worse than the  other c-APM agents. By reaching 2012-13 flu season,  suddenly GP agent performance improves compared to   other c-APM agents who  mainly use GFT data.  
 Given that GP agents had lower performance previously (in late   2011-12 flu season)  compared to other c-APM agents, c-APM relies more on other participating agents than GP agent. Once the performance of GP improves (in early  2012-13 flu seasons), GP   outperforms c-APM for a several number of weeks.  However, as shown by Figure~\ref{GP2008-2013}, c-APM recovers rapidly and adapts to the new conditions of the markets (i.e changing quality of market participants) in a few weeks.

 As mentioned earlier,~\cite{lampos2015advances} states  that that  2009-10 flu season    is a unique flu period since  none of the models experimented by them nor GFT could make prediction close to the ground truth (CDC data). Interestingly, c-APM achieves much less error than GP in 2009-10 as shown by Figure~\ref{GP-ErrorForeachPeriod}.

\begin{table}[]
\centering

\begin{tabular}{|c|c|c|c|c|}
\hline
\textbf{Period} & \textbf{Weeks} & \textbf{\begin{tabular}[c]{@{}c@{}}c-APM \\ MAE $\times 10^2$ \end{tabular}} & \textbf{\begin{tabular}[c]{@{}c@{}}GP \\ MAE $\times 10^2$ \end{tabular}} & \textbf{P-value} \\ \hline
2008-09         & 48             & 0.164                                                        & 0.175                                                      & 2.75E-01         \\ \hline
2009-10         & 57             & 0.156                                                        & 0.451                                                      & 7.40E-06         \\ \hline
2010-11         & 52             & 0.119                                                        & 0.130                                                      & 3.22E-01         \\ \hline
2011-12         & 52             & 0.098                                                        & 0.129                                                      & 6.31E-03         \\ \hline
2012-13         & 65             & 0.220                                                        & 0.198                                                      & 8.73E-01         \\ \hline
2008-2013       & 274            & 0.155                                                        & 0.221                                                      & 3.65E-05         \\ \hline
\end{tabular}
\caption{Performance of c-APM and GP in predicting ILI rate using Mean Absolute Error (MAE)  and p-values of paired t-test. }
\label{GpTable}
\end{table}

 \section {Analysis}\label{ssConclusion}

 c-APM   outperforms both the Google Flu Trend and GP models because: 
 \begin{enumerate}[i)]
 \item   c-APM integrates  different  data sources such as   CDC reports and Google Flu Trend prediction for different states and cities of USA. 
 \item   c-APM analyses each data source  with a variety of machine learning models and combines their results.
 
\item   c-APM adjusts the influence of  agents on market prediction  automatically according to their  quality. Over time, high quality agents -- either because of their effective analysis model or accessing high quality data source --  gain more revenue than low quality agents. Therefore, high quality agents accumulate more budget  and  they can make  larger investment on their prediction than poor performing agents. Subsequently, high quality agents  achieve   larger influence in the market   as the integration function weights each prediction by its corresponding   investment amount. 

 \item   c-APM adapts to the dynamic environment where the quality of   data sources and   the performance of a model on each data source fluctuates over time. Once the quality of an agent prediction changes, its performance in the market is affected and hence  the  influence of agents on the market prediction is tuned according to their current quality, as explained above.

\item The Q-learning trading strategy causes high quality agents lead the market by preserving their original prediction  and low quality agents follow them and hence minimise their negative effect   on forming  market prediction. 

 \item c-APM can minimise the effect of misleading factors and noise  since c-APM   integrates  various data sources  and combines  the result of different machine learning models, while dynamically changing their weight according to their varying quality. 
 For example, as shown by   Figure~\ref{errorCompareGFTall}, Google Flu Trend overestimated the Flu rate by large extent  in 2013 due to a misleading factor which is a  heightened  media coverage on the severity of the flue~\citep{blog.google10-29-2013}. 
  Since Google Flu Trend is being used as one of c-APM data sources,  c-APM also overestimates   the Flu rate to some extent  but much less than Google Flu Trend.        In c-APM, as soon as an agent  loses  its quality,  it either loses most of its budget or learns  to improve its prediction using wisdom of the crowd as advised by its Q-learning trading strategy. Therefore, its original influence on the market prediction decreases and c-APM relies to other agents with higher current performance. In the similar way, once the quality of an agent improves,  the influence of the agent on forming the market prediction increases.

 \end{enumerate}

  \bibliographystyle{abbrvnat}

\bibliography{references}
 
\end{document}